\documentclass[3p,numbers]{elsarticle}
\usepackage{amsmath}
\usepackage{amssymb}
\usepackage{nomencl}
\makenomenclature

\usepackage{graphicx}
\usepackage{bm}
\usepackage[usenames]{xcolor}
\usepackage{hyperref}
\usepackage{subcaption}
\usepackage{multicol} 

\begin{document}
\title{Background noise pushes azimuthal instabilities away from spinning states}

\author{G. Ghirardo\corref{cor1}}
\ead{giulio.ghirardo@ansaldoenergia.com}
\author{F. Gant}
\address{Ansaldo Energia Switzerland, R\"omerstrasse 36, Baden 5401, CH}

\date{\today}
\begin{abstract}
Azimuthal instabilities occur in rotationally symmetric systems, either as spinning (rotating) waves or standing waves. We make use of a novel ansatz to derive a differential equation characterizing the state of these instabilities in terms of their amplitude, orientation, nature (standing/spinning) and temporal phase. For the first time we show how source terms determine quantitatively the system preference for spinning and/or standing states. In particular we find that, when present, background noise pushes the system away from spinning states and towards standing states, consistently with experiments.
\end{abstract}
 
\maketitle

\vspace{-.3cm}
\section{Introduction}

Azimuthal instabilities occur as fluctuations of the pressure and velocity fields in geometries that exhibit a full rotational symmetry, like an axisymmetric wake behind a bluff body, or a discrete rotational symmetry, like an annular combustor. They can also occur as mechanical vibrations of a circular membrane, like a drum, or otherwise rotationally symmetric object, like a cymbal. The symmetry may not be exact for their occurrence, as is common in real-world applications where imperfections within the manufacturing tolerances are accepted. We derive the results for acoustics in the following, but the results hold for azimuthal instabilities in general.

We focus on systems where only one dominant peak at the frequency $\omega$ is present in the spectrum of the fluctuating field, plus other peaks at the multiples of $\omega$ due to the possibly nonlinear system behaviour. In other words, we neglect cases where more than one mode oscillate at the same time.

\begin{figure}
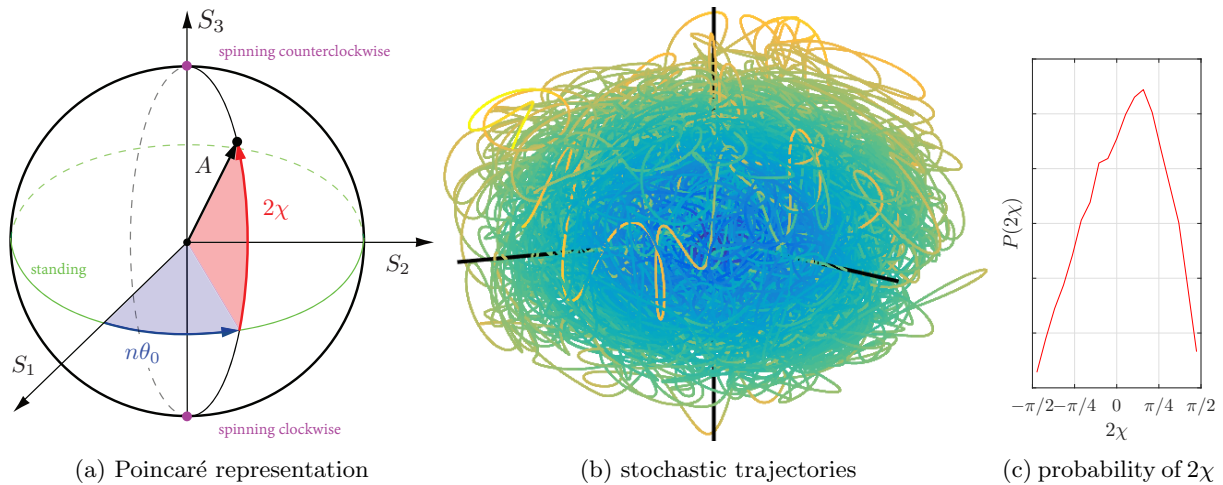

  \centering
  \begin{subfigure}[t]{.34\columnwidth}
    \centering\includegraphics[width=\textwidth]{{{fig4}}}
    \caption{Poincar\'e representation}
  \end{subfigure}
  \begin{subfigure}[t]{.44\columnwidth}
    \centering\includegraphics[width=\textwidth]{{{trajectories3}}}
    \caption{stochastic trajectories}
  \end{subfigure}
    \begin{subfigure}[t]{.17\columnwidth}
    \centering\includegraphics[width=\textwidth]{{{P2chi}}}
    \caption{probability of $2\chi$}
  \end{subfigure}
  \caption{a) Poincar\'e representation of the system state, characterized by the amplitude/radius $A$, the latitude angle $2\chi$ and the longitude angle $n\theta_0$. Points on the equator represent pure standing waves. Points at the north and south poles represent waves spinning respectively in the counterclockwise and clockwise direction. \label{FigSphere} In the same 3-dimensional space of a), we present in b) an example of the system state trajectory for the azimuthal instability of a nominally rotationally symmetric industrial annular combustor. The trajectory is described by the polar coordinates $(A(t),2\chi(t), n\theta_0(t))$ as function of time $t$ for approximately $130'000$ acoustic periods. The colour of the line describes qualitatively the amplitude $A$ of the point. c) probability density function of the angle $2\chi$ for the trajectory in b), showing that the system is never in spinning states at $\chi=\pm\pi/4$. This happens because of the background noise, which pushes the system away from the poles of the sphere in a). Data and figures were originally discussed in \cite{Ghirardo2018PRF}. }
\end{figure}

Recently Ghirardo and Bothien \cite{Ghirardo2018PRF} proposed a new ansatz for these fluctuations:
\begin{align}
\label{q}
p_1(\theta,t) =&A\cos(n(\theta-\theta_0))\cos(\chi)\cos(\omega t + \varphi) +A\sin(n(\theta-\theta_0))\sin(\chi)\sin(\omega t + \varphi)
\end{align}
In \eqref{q} $p_1(\theta,t)$ is the fluctuating pressure field, the positive integer $n$ is the azimuthal order of the instability and the four variables $\{A,n\theta_0,\chi,\varphi\}$ depend all on the time $t$. For a discussion of the advantages of \eqref{q} over the more conventional combinations of two counter-rotating spinning waves or two fixed standing waves, we refer the reader to \cite{Ghirardo2018PRF}. We only recall next the physical interpretation of the variables $\{A,n\theta_0,\chi,\varphi\}$. $A$ is the non-negative amplitude of the solution. The orientation angle $n\theta_0$ is the location of the pressure antinode of the standing component of the acoustic field. The angle $\chi$ is bounded in $[-\pi/4,\pi/4]$ and is called the nature angle because it quantifies the nature of the solution, i.e.~whether the system is standing ($\chi=0$), spinning ($\chi=\pm\pi/4$), or in a state between standing and spinning. $\varphi$ is the temporal phase of the oscillation. These four state space variables $\{A,\chi,n\theta_0,\varphi\}$ can be used in low order models and can be reconstructed from numerical/experimental time series, allowing a direct validation of theoretical results. The three variables $\{A,n\theta_0,2\chi\}$ can be interpreted as polar coordinates on the Poincar\'e sphere, as presented in Fig.~\ref{FigSphere} \cite{Ghirardo2018PRF}.

Many applications are noisy environments. For example, experiments show that annular combustors exhibit a high level of background noise, as discussed for example in Fig.~\ref{FigSphere}.b for an industrial engine and reviewed later in \S\ref{sexper}. No theoretical nor experimental study to date has discussed the effect of the intensity $\sigma$ of the background noise on the nature angle $\chi$ of the solution, i.e.~on  whether the system prefers spinning or standing solutions. The ansatz \eqref{q}, where the nature angle $\chi$ is a state space variable of the problem, offers this opportunity. It will be sufficient to characterize the dynamical behaviour of $\chi$ as function of the noise intensity $\sigma$.

The paper layout is as follows. In \S\ref{stheory} the ansatz \eqref{q} is substituted into the governing equations, the dynamical system is characterized, and general predictions are drawn. In \S\ref{sexper} a comparison with experimental evidence is presented, with good qualitative agreement with the predictions. We draw the conclusions in \S\ref{sconcl}.

\vspace{-.2cm}
\section{Theory}
\label{stheory}
The governing equations are the fluctuating pressure and momentum equations, assuming inviscid flow, linear acoustics, zero mean flow, as reviewed e.g.~by Clavin et al.~\cite[Appendix \S A]{Clavin94}:
\begin{subequations} 
\label{fe2f2e2}
\begin{align}
\label{eq_1}
\frac{1}{c^2}\frac{\partial p_1}{\partial t} + \rho_0\nabla\cdot{\bf{u}}_1&=\frac{\gamma-1}{c^2}q_1 +\frac{\sigma}{c^2}\xi\\
\label{eq_2}
\frac{\partial {\bf{u}_1}}{\partial t}+\frac{1}{\rho_0}\nabla p_1&=0
\end{align}
\end{subequations}
where $\rho_0$ is the mean density, $p_1$ the fluctuating pressure field, ${\bf{u}_1}$ the fluctuating velocity field and $c$ is the speed of sound. The right hand side of \eqref{eq_1} is a generic source term, in its deterministic component $q_1$ and its stochastic component $\sigma\xi$. $\xi(\theta,t)$ is a stochastic field and $\sigma$ is its noise intensity. In thermoacoustic applications the deterministic source term models the fluctuating heat release rate minus the acoustic damping, and the stochastic component occurs due to turbulent fluctuations at the flame front \cite{Rajaram2009}. We model the stochastic component as an additive and delta correlated noise process, and assume that $\xi$ does not depend on the azimuthal angle $\theta$, so that
\begin{align}
\label{pippero}
\frac{1}{\pi}\int_0^{2\pi} e^{in(\theta-\theta_0)}\xi(\theta,t) d\theta= \xi_1(t)+i\xi_2(t),
\end{align}
where $\xi_1(t)$ and $\xi_2(t)$ are independent, white gaussian sources with unit variance.

One studies the problem \eqref{fe2f2e2} in cylindrical coordinates. By direct substitution of \eqref{q} into the azimuthal component of \eqref{eq_2} the solution for the azimuthal acoustic velocity $u_{1,\theta}$ is recovered. One then substitutes $u_{1,\theta}$ and \eqref{q} into the azimuthal component of \eqref{eq_1} and applies the method of stochastic averaging. This method treats the system as weakly nonlinear, assuming the right hand side of \eqref{eq_1} is small, and averages the response of the system over one period $2\pi/\omega$, obtaining a new set of equations in terms of $\{A,n\theta_0,\chi,\varphi\}$:
\begin{align}
\nonumber
(\ln A)'+\left(n\theta_0'+\varphi'\sin(2\chi)\right)&i+\varphi'\cos(2\chi)j-\chi'k = \frac{1}{2}\frac{1}{2\pi}\int_0^{2\pi} \left(e^{i2n(\theta-\theta_0)}e^{k\chi}+e^{-k\chi}\right)Q_\theta(A,\chi,n\theta-n\theta_0)d\theta\,e^{k\chi}\\
\label{bb}
&+\left(-\frac{\omega}{2}+\frac{\omega_0^2}{2\omega}\right)e^{-k\chi}je^{k\chi}+\frac{\sigma^2}{4A^2}\left(1+\tan(2\chi)k\right)+\frac{\sigma}{\sqrt{2}A}\mu_z
\end{align}
The resulting equation \eqref{bb} is a quaternion-valued differential equation. Quaternion numbers comprise of three imaginary units $i,j,k$ instead of one imaginary unit $i$ as in the case of complex numbers. The three units $i,j,k$ are defined by $i^2=j^2=k^2=-1$ and $ij=k$, $jk=i$, $ki=j$. A generic quaternion number can be written as $z=z_0+iz_1+jz_2+kz_3$, where $z_0,z_1,z_2,z_3$ are real numbers.

In \eqref{bb}, $Q_\theta$ is the describing function \cite{Gelb68} of the deterministic source term $(\gamma -1)q_1$ appearing in \eqref{fe2f2e2}, which may depend also directly on $\theta$. We leave to a later work a detailed derivation and discussion of \eqref{bb}, since a comprehensive evaluation is not needed for a Rapid Communication. We focus on the $k$-imaginary component of the system \eqref{bb}:
\begin{align}
\label{eqchichichi2}
\chi'&=\left[\frac{1}{2}\frac{1}{2\pi}\int_0^{2\pi} \left(e^{i2n(\theta-\theta_0)}e^{k\chi}+e^{-k\chi}\right)Q_\theta(A,\chi,n\theta-n\theta_0)d\theta\,e^{k\chi}\right]_k-\frac{\sigma^2}{4 A^2}\tan(2\chi)+\frac{\sigma}{\sqrt{2}A}\mu_3
\end{align}
where the subscript on the square bracket denotes the $k$-imaginary component of the term between brackets, and $\mu_3$ is a white gaussian noise term. To the knowledge of the authors, this is the first time that the nature of the solution $\chi$ (standing vs spinning) appears in a differential equation.

In the deterministic case $\sigma$ is zero and the $k$-imaginary component of the term between square brackets is the only term at the right hand side of \eqref{eqchichichi2}. It determines the nature of the solution, i.e.~whether standing ($\chi=0$) and spinning ($\chi=\pm\pi/4$) states are fixed points of \eqref{eqchichichi2} and if they are stable or not. This component describes the effect of explicit symmetry breaking, where $Q$ directly depends on $\theta$, and spontaneous symmetry breaking, which occurs due to the term $e^{i2n(\theta-\theta_0)}$. A discussion of the effect of spontaneous symmetry breaking in this deterministic case is reviewed later in \S\ref{sexper}.

In the stochastic case, i.e.~when the level $\sigma$ of the background noise is not negligible, the last two terms in \eqref{eqchichichi2} are not zero. Both of them are functions of the ratio $\sigma/A$ and are discussed next. The second to last term $-\sigma^2\tan(2\chi)/4 A^2$ tends to $\mp\infty$ for $\chi\rightarrow\pm\pi/4$, and then spinning states $\chi=\pm\pi/4$ cannot be fixed points of the system in the stochastic case. The noise pushes the system state towards positive values when $\chi$ is negative, and towards negative values when $\chi$ is positive. In other words, the noise pushes the state variable $\chi$ away from the boundaries $\pm\pi/4$ of the domain for $\chi$, which are spinning states, towards standing states $\chi=0$.

The last term at the right hand of \eqref{eqchichichi2} is large in absolute value at small amplitudes, at which the nature angle of the system can change very quickly. This is expected, since when the pressure field is really small a little perturbation can completely change the solution, and hence the nature angle $\chi$. However, as the system state approaches the spinning states at $\chi=\pm\pi/4$, the second-to-last term becomes more dominant, because it scales like $A^{-2}$. It guarantees that the variable $\chi$ stays in the bounded domain $(-\pi/4,\pi/4)$.
 
\vspace{-.2cm}
\section{Comparison with experiments}
\label{sexper}
In this section we discuss how the theory compares with available experimental and engine data of nominally rotationally symmetric combustors, and for which a discussion of the standing and spinning nature of the system is available. Experiments that only change the level of background noise $\sigma$ and keep all other relevant parameters the same have not yet been conducted. However, we can discuss if experiments of annular combustors are noisy or not, and if they manifest a statistical preference for standing or spinning states.

To start with, we review the analysis of combustors exhibiting a low level of background noise. To the knowledge of the authors, this regards only the MICCA combustor equipped with matrix burners \cite{Bourgouin2015}. This experiment exhibits low level of background noise, based on the fact that the system can converge to spinning or standing states and lingers in the vicinity of these solution with very tiny variations. In this case, one can assume that the effect of the background noise is negligible and treat the problem in a deterministic setting, i.e.~by setting $\sigma$ to zero in the equations here discussed. Theoretical conditions for stable limit-cycle solutions in a deterministic framework have been proposed by \cite{Ghirardo16JFM}, ignoring the effect of transverse forcing on the flames. They have been validated on experimental results of the MICCA combustor in this configuration for spinning modes in the same reference, and for standing modes by Laera et al.~\cite{Laera2017a}. Summarizing, at negigible levels of background noise both standing and spinning states are observed in experiments and predicted to be stable periodic solutions.

We turn our attention to the available cases of rotationally symmetric annular combustors exhibiting a non-negligible level $\sigma$ of background noise. In the annular rig of Worth et al.~\cite[Fig.~8]{Worth2013modaldyn}, modes are never purely spinning, but always between spinning and standing states, or primarily standing. The level of background noise of this combustor is not negligible, and the system keeps switching between states in a stochastic manner. In the MICCA combustor equipped with swirler-stabilized flames \cite[Fig.~14]{Bourgouin2013_asme}, the system state is dominantly standing, with a slight preference for mixed states between standing and spinning counter-clockwise. Also in this case the level of background noise is not negligible, because the PDF of the spin ratio in their Fig.~14 is heavy tailed. 
The industrial annular combustor discussed by Ghirardo et al.~ \cite[Fig. 6, before dampers' installation]{Ghirardo2018PRF}, and here reported in Fig.\ref{FigSphere}.c, shows a system state that is never fully spinning. Also this combustor shows a non negligible level of noise, as depicted by the trajectories of the system state on the Poincar\'e sphere in Fig.~\ref{FigSphere}.b. All this evidence shows that annular combustors exhibiting non-negligible levels of background noise are never in a spinning state. 
\vspace{-.2cm}
\section{Conclusions}
\label{sconcl}
We make use of a novel ansatz into the governing equations of azimuthal instabilities. We apply the method of averaging and obtain a novel quaternion-valued differential equation that governs the dynamics of the system in terms of state space variables that allow an intuitive physical interpretation. We predict that the background noise pushes the system away from spinning solutions, towards standing solutions. We show how this is consistent with existing experimental results. In particular, only experiments subject to a negligible level of noise exhibit purely spinning solutions. Conversely, experiments and combustors that are noisy never experience purely spinning solutions, but either states between spinning and standing solutions or mostly standing solutions. A future paper in this same journal will present a detailed account of the derivation of the averaged equations and a detailed interpretation of the novel differential equation. 

\bibliographystyle{elsarticle-num}
\bibliography{/users/axe/library}
\end{document}